\documentclass[showpacs,preprint,amsmath,amssymb]{revtex4}

\usepackage{graphicx}
\usepackage{epsfig}
\usepackage{xcolor}
\usepackage{dcolumn}
\usepackage{longtable}
\usepackage{longtable}

\begin{document}

\title{Quantum-mechanical description of angular motion of fission fragments at scission}


\author{T. M. Shneidman}
\affiliation{Joint Institute for Nuclear Research, Dubna, 141980, Russia}
\author{A. Rahmatinejad}
\affiliation{Joint Institute for Nuclear Research, Dubna, 141980, Russia}
\author{G. G. Adamian}
\affiliation{Joint Institute for Nuclear Research, Dubna, 141980, Russia}
\author{N. V. Antonenko}
\affiliation{Joint Institute for Nuclear Research, Dubna, 141980, Russia}

\date{\today}

\begin{abstract}
The quantum-mechanical description of the collective angular motion in a system  of  two touching fission fragments is proposed. The main peculiarities of excitation spectrum and the structure of the wave functions are investigated.
As found, the angular motion approximately corresponds to independent vibrations of  fragments  around the pole-to-pole configuration.
The model allows us to explain the experimentally observed lack of correlation between the angular momenta of fission fragments. Additionally the correlation between angular momentum and fragment mass is primarily linked to the change of fragments deformation. The saw-tooth behavior of angular momentum distribution with respect to the fragment mass is well explained.
\end{abstract}

\pacs{}

\maketitle

\section{Introduction}

After crossing the fission barrier, the fissioning nucleus goes into a scission configuration in which it can be treated as a dinuclear system (DNS) -- system of two deformed fragments in touching. Before decay, the DNS evolves by changing masses, charges and deformations of the fragments. This evolution is crucial for the formation of charge, mass, total kinetic energy and neutron multiplicity distributions of fission products. Additionally, the DNS fragments undergo angular collective motion as, for example, small-amplitude vibrations of the fragments around pole-to-pole configurations, which corresponds to the minimum of potential energy. This motion leads to striking phenomenon that even for spontaneous fission of even-even nuclei the fission fragments have rather large angular momentum \cite{Wilhelmy1972,Pleasonton1972,Wolf1976}. Experimentally, the angular momenta of fission fragments were determined in coincidence  of $E2$-transitions between yrast states of partner fragments. The experiments on the study of spontaneous fission of $^{252}$Cf performed using GAMMASPHERE detector provided  information about the correlation between  average angular momenta  of fission fragments and  number of neutrons emitted \cite{Akopian1994,Akopian1996,Akopian1997,Musangu2020}.

For the first time, theoretical analysis of different angular vibrational modes at scission leading to generation of angular momentum were performed in Ref.~\cite{Nix1965}.
This classical analysis was later supplemented by the quantum-mechanical treatment of the system consisting of deformed and spherical fragments \cite{Rasmussen1969}, and of the bending vibrational mode in the system of two deformed fragments \cite{Zielinska1974}. The bending and wriggling modes were further studied in \cite{Hess1984,Misicu1997,Shneidman2003,Kadmensky2015}.

Renewed interest to the problem of angular momentum generation is related to the recent experimental observations \cite{Wilson2021}, which revealed new remarkable information. Firstly, the angular correlation  with different constraints imposed on one of the fragments have demonstrated the almost absence of correlations between the angular momenta of fission fragments. Secondly, it was observed that the dependence of average spins on fragments mass demonstrates a saw-tooth behavior. A variety of theoretical models have been recently proposed to address the new experimental observation including the statistical approach \cite{Randrup2021,Randrup2022,Dossing2024}, DFT and TD-DFT with projection \cite{Bertsch2019,Marevic2021,Bulgac2022}, and the models based on time dependent dynamics with the collective Hamiltonian of the angular vibrational modes \cite{Scamps2023,Scamps2024}. Generally, most of  approaches agree on the role of collective angular motion in generation of  angular momenta of fission fragments.
Calculations in various models and with different sets of initial assumptions have shown a reasonable description of existing experimental data at least on a qualitative level. This suggests that  most  peculiarities of the generation of angular momenta of fission fragments are deeply based on the characteristic of collective mode itself rather than on a particular model used.

In order to analyze the main features of the collective angular motion, we present simple but general quantum-mechanical description of collective motion in DNS. We investigate the main characteristics of the angular vibrations which follows directly from the structure of the collective Hamiltonian. It will be shown that most of  experimental observations including saw-tooth behavior of average angular momentum versus fragment mass distribution and absence of correlation between the spins of fission fragments can be understood directly from our quantum consideration.

\section{Model}
At scission, the fissile nucleus is considered as a DNS consisting of two nuclei with masses $A_{1,2}$, charges $Z_{1,2}$, and quadrupole deformation parameters $\beta_{1,2}$. Here, $A_{1}+A_{2}=A$, and $Z_{1}+Z_{2}=Z$, where $Z$ and $A$ are the charge and mass of the fissile nucleus, respectively. We assume that the fragments are axially-symmetric. The vector ${\bf R}$ connects the centers of masses of the fragments.

\begin{figure}
    \centering
    \includegraphics[width=0.9\linewidth]{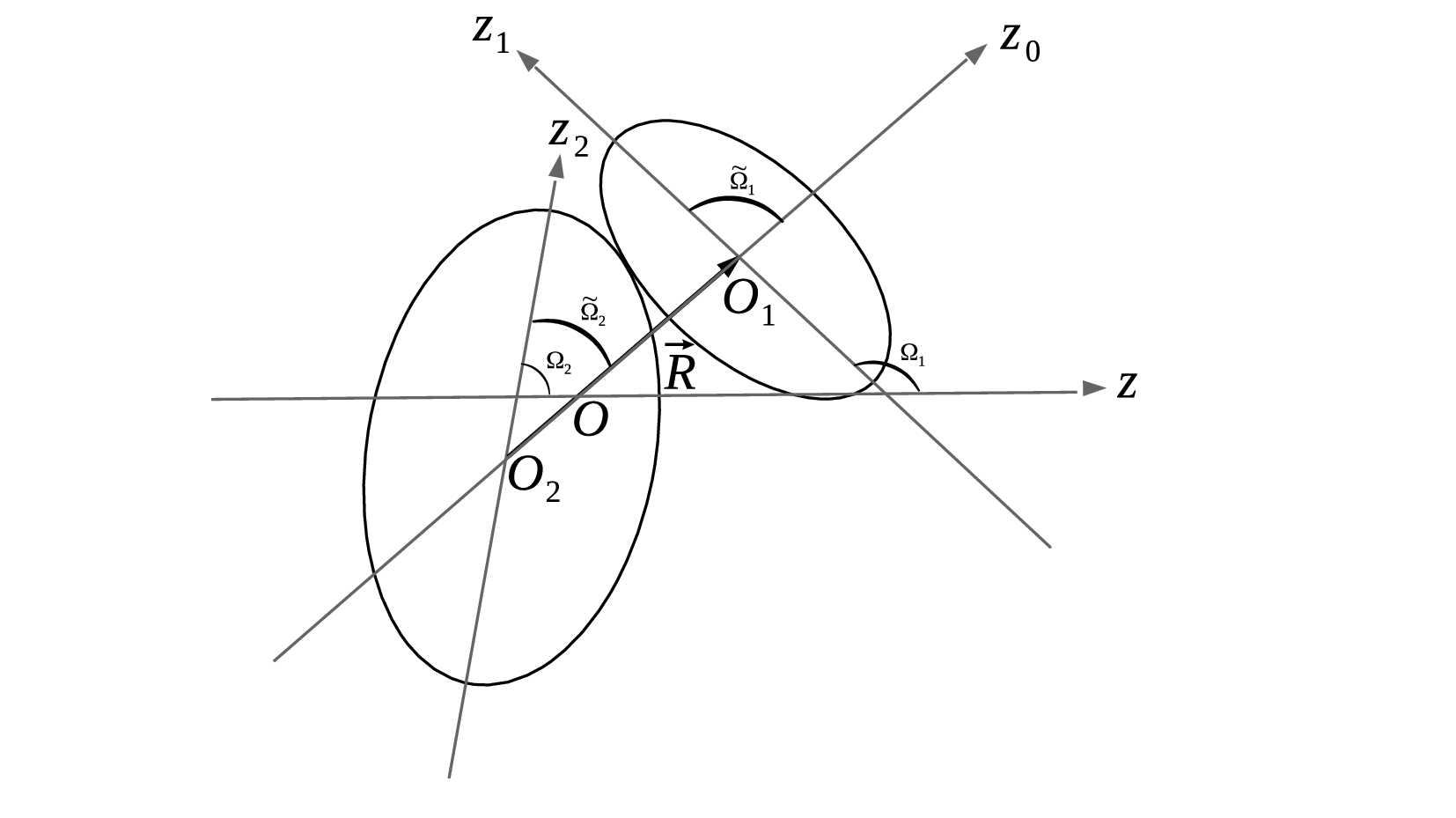}
    \caption{Schematic picture of the DNS. The laboratory system is denoted as $Oz$, where the center of laboratory system $O$ coincides with the center of mass of DNS. The body-fixed system $Oz_{0}$ is chosen so that its $z$-axis coincides with vector of relative distance ${\bf R}$. The orientation of body-fixed system with respect to laboratory system is defined by the Euler angles $\Omega_{0}=(\phi_{0},\theta_{0},0)$. The orientations of intrinsic systems of the fragments $O_1z_1$ and $O_2z_2$ with respect to the laboratory system are $\Omega_{1,2}=(\phi_{1,2},\theta_{1,2},0)$ and with respect to the body-fixed system are $\tilde{\Omega}_{1,2}=(\gamma_{1,2},\varepsilon_{1,2},0)$.}
    \label{schematic}
\end{figure}

To describe the orientation of each fragment as well as the DNS as a whole with respect to the laboratory coordinate system, we define three sets of the Euler angles (see the schematic picture in Fig.\ref{schematic}). Two sets, $\Omega_{1}=(\phi_{1},\theta_{1},0)$ and $\Omega_{2}=(\phi_{2},\theta_{2},0)$,  are the Euler angles  for intrinsic coordinate systems of corresponding fragments. These systems are chosen in such a way that their $z$-axes coincide with symmetry axes of the  fragments. The third angle of the set is redundant and  taken as zero because of the axial symmetry of the fragments.  Additionally, we introduce the Euler angles $\Omega_{0}=(\phi_{0},\theta_{0},0)$ which define the body-fixed system whose $z$-axis aligns with the direction of ${\bf R}$.

The Hamiltonian of the DNS at fixed mass (charge) asymmetry is written as
\begin{eqnarray} \label{eq1}
H=\frac{\hbar^2\hat{I}_{0}^2}{2\mu R^2}+\frac{\hbar^2\hat{I}_{1}^2}{2\Im_{1}}+\frac{\hbar^2\hat{I}_{2}^2}{2\Im_{2}}+V(R,\Omega_{0},\Omega_{1},\Omega_{2}).
\end{eqnarray}
The kinetic energy describes the rotation of the fragments and the rotation of DNS as a whole with respect to the laboratory system.
In Eq.~\eqref{eq1}, $\Im_{1,2}$, and $\Im_{R}=\mu R^2$  are the moments of inertia of the fragments and  DNS as a whole, respectively. The reduced mass of the DNS is $\mu=m_{0}A_{1}A_{2}/A$, where $m_{0}$ is the nucleon mass. The angular momentum operators are explicitly written in terms of Euler angles as
\begin{eqnarray} \label{eq2}
\hat{I}_{i}^2=-\left(\frac{1}{\sin\theta_{i}}\frac{\partial}{\partial\theta_{i}}\sin \theta_{i}\frac{\partial}{\partial\theta_{i}}+\frac{1}{\sin^2\theta_{i}}\frac{\partial^2}{\partial\phi_{i}^2}\right),\quad (i=0,1,2).
\end{eqnarray}

The eigenstates of kinetic energy operator are written in terms of tripolar spherical harmonics
\begin{eqnarray} \label{eq3}
\left[ i_{0}\times [i_{1}\times i_{2}]_{i_{12}} \right]_{(I,M)}&\equiv&
\left[Y_{i_{0}}(\Omega_{0})\times \left[Y_{i_{1}}(\Omega_{1})\times Y_{i_{2}}(\Omega_{2}) \right]_{i_{12}}\right]_{(I,M)}\nonumber \\
&=&\sum
_{m_{0}m_{1}m_{2}m_{12}} C_{i_{0}m_{0},i_{12}m_{12}}^{IM}
C_{i_{1}m_{1},i_{2}m_{2}}^{i_{12}m_{12}}Y_{i_{0}m_{0}}(\Omega_{R})Y_{i_{1}m_{1}}(\Omega_{1})Y_{i_{2}m_{2}}(\Omega_{2}).
\end{eqnarray}

Here, the angular momentum of the system as a whole $i_{0}$, and the angular momenta of fragments $i_{1}$ and $i_{2}$ are coupled to the state with total angular momentum $I$ and its projection $M$. For the case of even-even spontaneously fissioning nucleus considered here $I=M=0$ which leads to $i_{12}=i_0$.

In order to find  the potential energy $V$, we take into account that it does not depend on orientation of the system as a whole. Therefore, one can first calculate it with respect to the body-fixed frame, and then transform back into the laboratory coordinate system. Denoting the Euler angles describing the orientation of intrinsic systems of fragments with respect to body-fixed system as $\tilde{\Omega}_{1,2}=(\gamma_{1,2},\varepsilon_{1,2},0)$, the interaction potential $V$ is calculated as a sum of the Coulomb and nuclear interaction potentials  $V(R,\tilde{\Omega}_{1},\tilde{\Omega}_{2})=V_{N}(R,\tilde{\Omega}_{1},\tilde{\Omega}_{2})+V_{C}(R,\tilde{\Omega}_{1},\tilde{\Omega}_{2})$.

For nuclear part $V_N$ one can apply various approaches. In this study we use the double-folding potential as described in detail in Ref.~\cite{Adamian1996}. Our analysis showed that for any given orientation of fragments the angular dependence of the potential energy is mainly determined by the Coulomb interaction while the  nuclear interaction remains almost constant for touching fragments.Therefore, we assume
\begin{eqnarray} \label{eq11b}
V(\tilde{\Omega}_{1},\tilde{\Omega}_{2})=V_{C}(R_{m}(\tilde{\Omega}_{1},\tilde{\Omega}_{2}),\tilde{\Omega}_{1},\tilde{\Omega}_{2})+V_{N}(R_{m}(0,0),0,0),
\end{eqnarray}
where $R_{m}(\tilde{\Omega}_{1},\tilde{\Omega}_{2})$ is the distance corresponding to the touching configuration of the fragments for a given orientation and $V_{N}$ is taken for tip-tip orientation.
The Coulomb interaction is calculated in terms of multipole expansion as
\begin{eqnarray} \label{eq7}
V_{C}&=&\sum_{l_1l_2}V_{\lambda_0,\lambda_1,\lambda_2}(\beta_1,\beta_2)\left[Y_{\lambda_1}(\tilde{\Omega}_1)\times Y_{\lambda_2}(\tilde{\Omega}_2)\right]_{(\lambda_1+\lambda_2,0)}, \nonumber \\
V_{\lambda_0,\lambda_1,\lambda_2}&=&(-1)^{\lambda_2}\sqrt{\frac{(4\pi)^2(2\lambda_{1}+2\lambda_{2})!}{(2\lambda_{1}+1)!(2\lambda_{2}+1)!}} \frac{Q_{\lambda_1}^{(1)}(\beta_{1})Q_{\lambda_2}^{(2)}(\beta_{2})}{R_{m}^{\lambda_1+\lambda_2+1}(\tilde{\Omega}_{1},\tilde{\Omega}_2)},
\end{eqnarray}
where
\begin{eqnarray} \label{eq8}
Q_{\lambda}^{(i)}(\beta_{i})=\sqrt{\frac{4\pi}{2\lambda+1}}\int\rho_{i}(\textbf{r}_{i})r_{i}^{\lambda}Y_{\lambda 0}(\theta,\phi)d\textbf{r}_{i}
\end{eqnarray}
are multipole moments of the fragments $(i=1,2)$.
The potential energy \eqref{eq11b} is transformed to the laboratory system using the expression \cite{Varshalovich1988}
\begin{eqnarray} \label{eq11}
\left[Y_{\lambda_{1}}(\tilde{\Omega}_{1})\times Y_{\lambda_{2}}(\tilde{\Omega}_{2}) \right]_{(\lambda_{0}0)}=\sqrt{4\pi}\left[Y_{\lambda_{0}}(\Omega_{0})\times \left[Y_{\lambda_{1}}(\Omega_{1})\times Y_{\lambda_{2}}(\Omega_{2}) \right]_{\lambda_{0}}\right]_{(00)}.
\end{eqnarray}

The Hamiltonian describing the angular vibrations in DNS is as following
\begin{eqnarray} \label{eq12}
H&=&\frac{\hbar^2\hat{I}_{0}^2}{2\mu R_{m}^2(\tilde{\Omega}_{1},\tilde{\Omega}_{2})}+\frac{\hbar^2\hat{I}_{1}^2}{2\Im_{1}}+\frac{\hbar^2\hat{I}_{2}^2}{2\Im_{2}}\nonumber \\
&+&\sqrt{4\pi}\sum_{\lambda_{1},\lambda_{2}} V_{\lambda_{0},\lambda_{1},\lambda_{2}}\left[ (\lambda_{1}+\lambda_{2})\times [\lambda_{1}\times \lambda_{2}]_{(\lambda_{1}+\lambda_{2})} \right]_{(0,0)},
\end{eqnarray}
where the constant term $V_{N}(R_{m}(0,0),0,0)$ in the potential energy is omitted.
In order to simplify the diagonalization of $H$, we use the following expansion of inverse moment of inertia corresponding the whole system
\begin{eqnarray} \label{eq13}
\frac{1}{\mu R_{m}^{2}(\tilde{\Omega}_{1},\tilde{\Omega}_2)}=\frac{1}{\mu R_{m}^{2}(0,0)}\left[1+\sqrt{4\pi}\sum_{\lambda_{0},\lambda_{1},\lambda_{2}}R^{(2)}_{\lambda_{0},\lambda_{1},\lambda_{2}}\left[ \lambda_{0}\times [\lambda_{1}\times \lambda_{2}]_{\lambda_{0}} \right]_{(0,0)}\right].
\end{eqnarray}

The Hamiltonian \eqref{eq12} is diagonalized in a set of basis functions \eqref{eq3}. For the spontaneous fission of even-even nucleus, the eigenstates of $H$ are expressed as
\begin{eqnarray} \label{eq4b}
\psi_{n}=\sum_{i_{0}i_{1}i_{2}}a^{(n)}_{i_{0}i_{1}i_{2}}\left[i_{0}\times\left[i_{1}\times i_{2}\right]_{i_{0}}\right]_{(00)}.
\end{eqnarray}
Since fragments are assumed to be reflection-symmetric, the values of $i_{1,2}$ are even. Taking into account that parity of basis states \eqref{eq3} is given as $\pi=(-1)^{i_{1}+i_{2}+i_{0}}$, $i_{0}$ is also even and $|i_{1}-i_{2}|\leqslant i_{0}\leqslant|i_{1}+i_{2}|$.
For the state $\psi_{n}$, the value of $|a^{(n)}_{i_{0}i_{1}i_{2}}|^2$ gives the probability that the angular momentum of the first fragment is $i_{1}$, the second fragment is $i_{2}$ and the angular momentum of the system as a whole is $i_{0}$.

It is worth mentioning that for the case of $I^\pi=0^+$, the basis employed here has one to one correspondence with the basis states introduced in Ref.~\cite{Scamps2023}. Indeed,
\begin{eqnarray} \label{eq14}
&&\left[ Y_{i_0}(\Omega_{0})\times [Y_{i_1}(\Omega_{1})\times Y_{i_2}(\Omega_{2})]_{i_0} \right]_{(0,0)}=\frac{(-1)^{i_0}}{\sqrt{4\pi}}\sum_{k}C_{i_1k i_2-k}^{i_0 0}Y_{i_1 k}(\tilde{\Omega}_1)Y_{i_2 -k}(\tilde{\Omega}_2),\nonumber \\
&&Y_{i_1 k}(\tilde{\Omega}_1)Y_{i_2 -k}(\tilde{\Omega}_2)=\sum_{i_0}(-1)^{-i_0}\sqrt{4\pi}C_{i_1k i_2-k}^{i_0 0}\left[ Y_{i_0}(\Omega_{0})\times [Y_{i_1}(\Omega_{1})\times Y_{i_2}(\Omega_{2})]_{i_0} \right]_{(0,0)},
\end{eqnarray}
where $Y_{i_1 k}(\tilde{\Omega}_1)Y_{i_2 -k}(\tilde{\Omega}_2)$ up to a constant coincides with the basis used in Ref.~\cite{Scamps2023}.

It is interesting to analyze the basis states \eqref{eq3} in terms of known vibrational modes \cite{Nix1965}. For example, the states $\left[0\times\left[i\times i\right]_{0}\right]_{(00)}$, and $\left[2i\times\left[i\times i\right]_{2i}\right]_{(00)}$ represent pure bending and wriggling types of motion, respectively. Analogously, the states $\left[i\times\left[i\times 0\right]_{i}\right]_{(00)}$, and $\left[i\times\left[0\times i\right]_{i}\right]_{(00)}$, correspond to the modes when only one fragment is excited. However, since the potential energy mixes the various basis states in the wave functions, none of these modes can be purely attributed to a given state in which they have certain weights.

\section{Results and discussions}

\begin{figure}[t]
    \centering
   \fbox{ \includegraphics[width=0.3\linewidth]{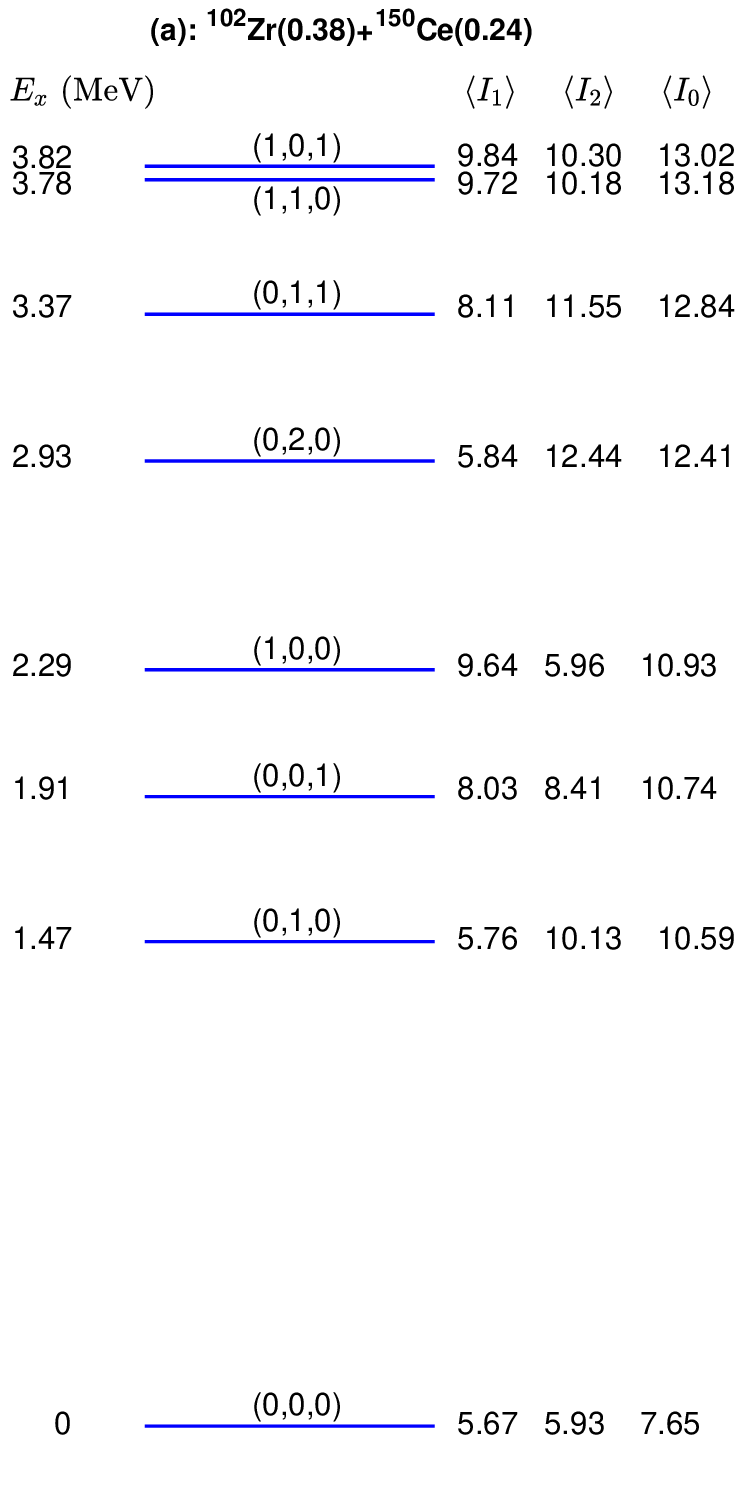}}
   \fbox{ \includegraphics[width=0.3\linewidth]{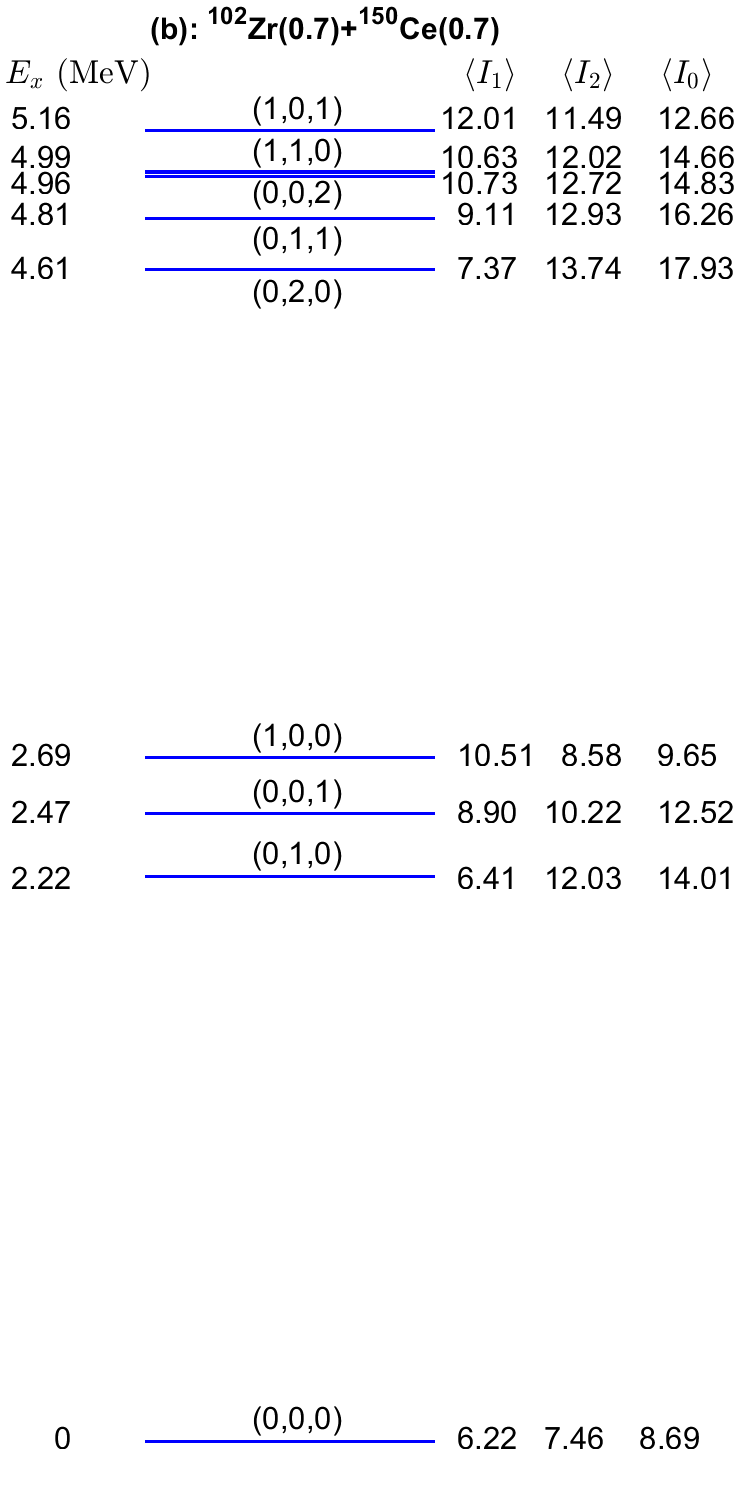}}
   \fbox{ \includegraphics[width=0.3\linewidth]{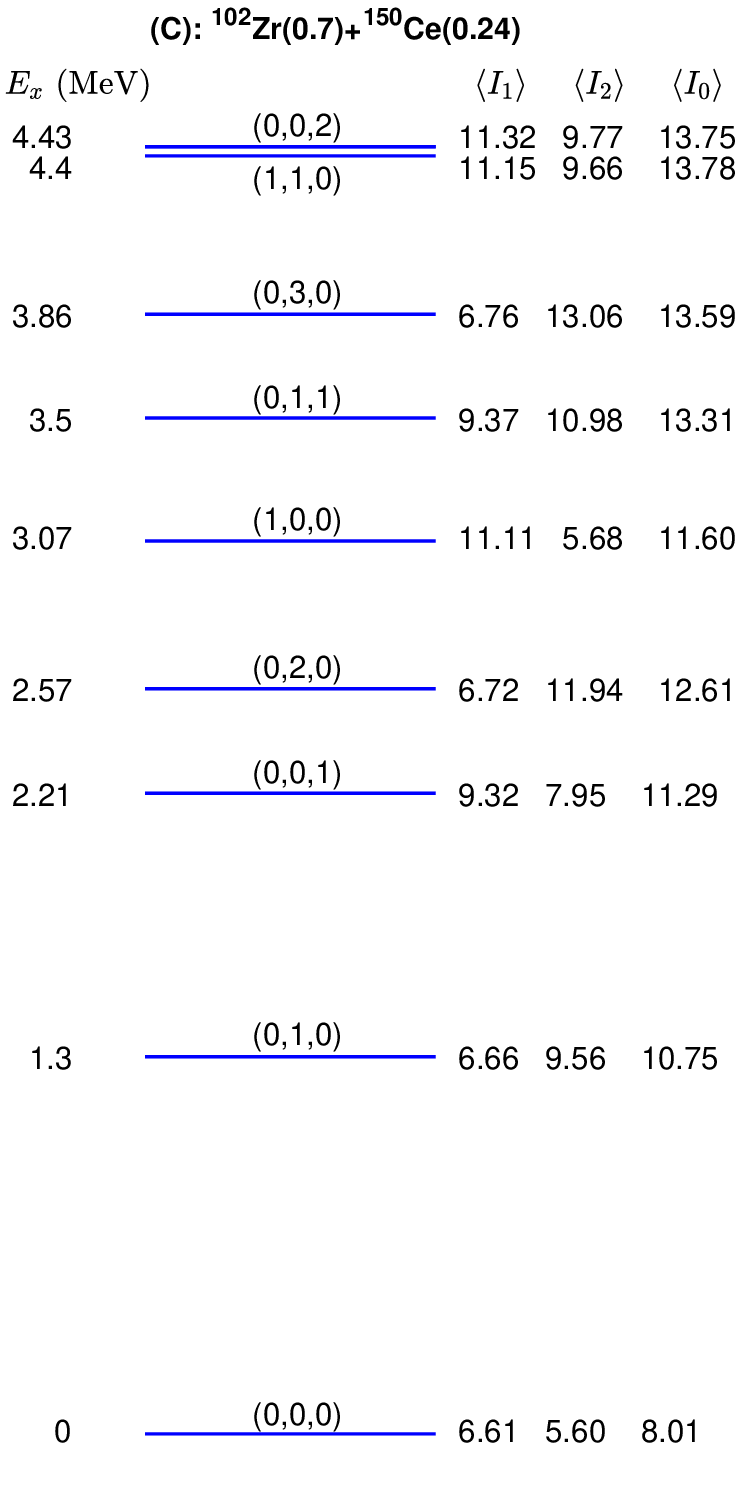}}
    \caption{Excitation spectra for the angular vibrations in $^{102}$Zr$(\beta_{Zr})$+$^{150}$Ce$(\beta_{Ce})\rightarrow ^{252}$Cf at different deformations $(\beta_{Zr},\beta_{Ce})$ as indicated. The numbers to the left of the lines give the energies, while the numbers to the right side denote the average angular momenta of the first fragment $\langle I_1 \rangle$, of the second fragment $\langle I_2 \rangle $ and of the rotation of DNS as a whole $\langle I_0 \rangle$. The approximate quantum numbers defined by Eq.~\eqref{approx} for each excited state are marked along the lines.}
    \label{Spec}
\end{figure}

In order to analyze the excitation spectra generated by the Hamiltonian \eqref{eq12}, we performed the calculations for the spontaneous fission of $^{252}$Cf into the primary fragments $^{102}$Zr+$^{150}$Ce for three sets of deformation parameters: $(\beta_{Zr},\beta_{Ce})=(0.32,0.2)$, $(0.7,0.7)$, and $(0.7,0.2)$.
The energies $E_x$ of excited states are listed in Fig.~\ref{Spec}. as well as the average angular momenta of the first fragment $\langle I_1 \rangle$,  second fragment $\langle I_2 \rangle$, and of the rotation as a whole $\langle I_0 \rangle$.
As seen, the average angular momenta of fragments increase with their deformation.

The spectra presented in Figs.~\ref{Spec}~(a,b) can be decomposed into the groups of states lying close to each other, thus, forming some  shells. To understand this structure, one should bear in mind that due to the stiff potential, the angular motion in DNS can be approximately treated as independent small-amplitude oscillations of the fragments around the pole-to-pole configuration. Neglecting the coupling between the angular vibrations of the fragments in this case the excitation spectrum of the angular motion is approximately written as
\begin{eqnarray}\label{approx}
E_{n_1,n_2,\bar{K}}=\hbar \omega_1 (2 n_1+|\bar{K}|+1)+\hbar \omega_2 (2 n_2+|\bar{K}|+1).
\end{eqnarray}
Here, the frequencies describing the oscillations of the fragments are
\begin{eqnarray}\label{Im12}
\omega_{i}=\sqrt{C_{i}/\Im_{b,i}},\quad (i=1,2),
\end{eqnarray}
where $C_i$ are the stiffness of the potential energy with respect to the $\varepsilon_i$ and the moments of inertia are $\Im_{b,i}=(1/\Im_0+1/\Im_i)^{-1} \approx \Im_i$ \cite{Shneidman2015}. In Eq.~\eqref{approx}, $n_{1,2}$ are the numbers of oscillator quanta and $K$ describes the projection of angular momentum of the fragments onto the direction of fragments separation ${\bf R}$, which is generated by the azimuthal rotations. Since  azimuthal rotation of one fragment can only be compensated by azimuthal rotation of another fragment: $K_1=-K_2=\bar{K}$.

Using Eq.~\eqref{approx}, one can associate the ground states with $n_1=n_2=\bar{K}=0$.
For the spectra presented in Fig.~\ref{Spec}(a,b), the states forming the shells correspond to the same numbers of oscillator quanta.
Indeed, the first group consists of states with energies $E_{1,0,0}=2\hbar \omega_1$, $E_{0,1,0}=2\hbar \omega_2$, and  $E_{0,0,1}=\hbar (\omega_1+\omega_2)$. The next shell consists of states characterized by two oscillator quanta and so on. As seen in Fig.~\ref{Spec}, the lowest excited angular vibrational state lies at the energy larger than 1 MeV as in Ref.~\cite{Shneidman2003}. Therefore, the ground state decisively contributes to the generation of angular momentum.

When the deformation of fragments are close to each other and $\omega_1 \approx \omega_2$, the excited states with the same number of quanta bundle together forming distinct shells (Fig.~\ref{Spec}(a) and (b)). If the deformations are quite different, the states are still described by the approximate expression \eqref{approx}, that is easily verified in Fig.~\ref{Spec}(c). However, the shells become intertwined due to significantly different frequencies.

As seen in Fig.~\ref{Spec}, the picture of independent vibrations of the fragments around pole-to-pole configuration seems to be suitable to explain the generation of angular motion in the fissioning system at scission.
If the concept of independent vibrations holds true, one can draw two main conclusions. Firstly, the distribution of $K$ values for any particular low-lying  state should be  peaked around corresponding approximate value  $\bar{K}=0,1,2...$ characterizing this state. Secondly, the angular momenta of the fragments at scission should be uncorrelated. By analyzing the wave functions of the low-lying states related to angular motion, we examine these assessments.

To construct the distribution of $K$ values, we use expression \eqref{eq14}. If the basis state $\left[i_{0}\times\left[i_{1}\times i_{2}\right]_{i_0}\right]_{(00)}$ contributes to the  wave function, it generates the particular $K$ value with probability $|C_{i_1K i_2-K}^{i_0 0}|^2$. Therefore, using Eq.~\eqref{eq4b} for the wave function,  the distribution of $K$ value as following
\begin{eqnarray} \label{Kdist}
P_n(K)=2 \sum_{i_{0}i_{1}i_{2}}|a^{(n)}_{i_{0}i_{1}i_{2}}|^2 |C_{i_1K i_2-K}^{i_0 0}|^2.
\end{eqnarray}

\begin{figure}[t]
    \centering
    \includegraphics[width=1\linewidth]{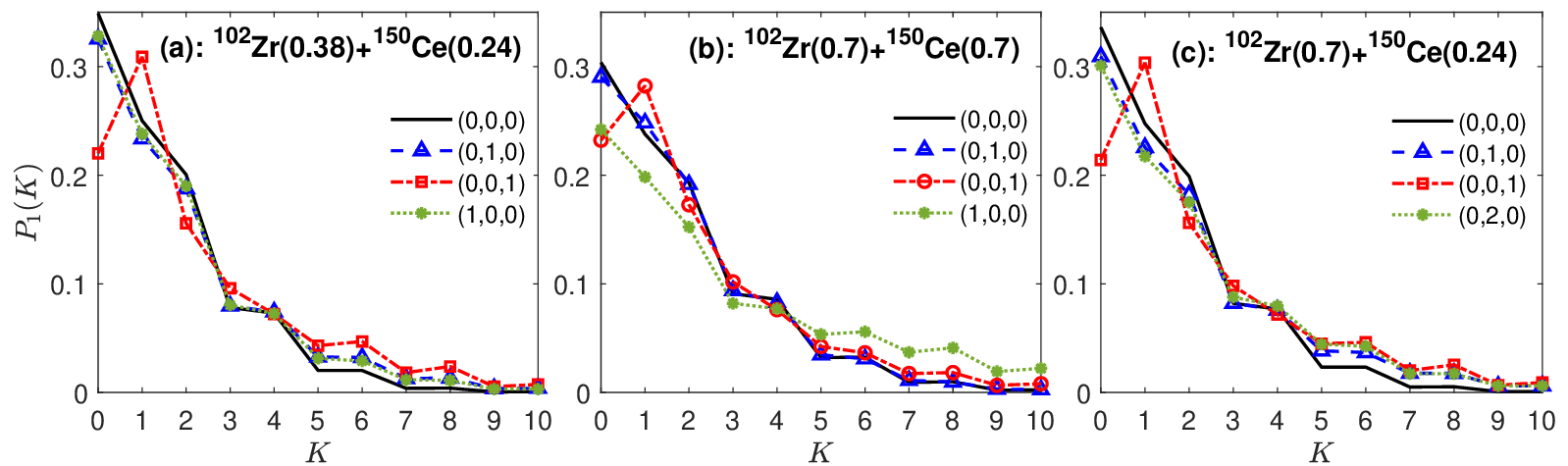}
    \caption{Probability distribution of $K$ values calculated using Eq.~\eqref{Kdist} for the lowest states of angular vibrations in $^{102}$Zr+$^{150}$Ce at different deformations of fragments. The excited states are marked with their approximate quantum numbers.}
    \label{K-distribution}
\end{figure}

The results of calculations for the lowest states of DNS $^{102}$Zr+$^{150}$Ce  for different deformations of the fragments are shown on Fig.~\ref{K-distribution}. Distributions of $K$ values shown in Fig.~\ref{K-distribution} are indeed peaked either around $K=0$ or $K=1$, but show  significant contribution of $K\leqslant 4$. One can note from Fig.~\ref{K-distribution}, that the distributions get wider with increasing deformation. As shown in Ref.~\cite{Shneidman2015} for the system consisting of spherical and deformed fragments, due to the coupling between vibration of deformed fragment and rotation of DNS as a whole, $K$ can be treated as a quantum number only approximately.

To investigate  correlations of angular momenta of the fragments, we adopt an approach which is identical to that in Ref.~\cite{Wilson2021}.  We calculate average angular momentum of the first fragment with a constraint that the angular momentum of the second fragment is larger than $I_{min}=0\hbar$, $2\hbar$, $4\hbar$, etc. This can be done by choosing corresponding amplitudes in the wave function \eqref{eq4b} as
\begin{eqnarray} \label{eq11ss}
\langle I_{1}\rangle_{n}=\left[\sum_{i_{0},i_{1},i_{2}\geqslant I_{min}}|a^{(n)}_{i_{0}i_{1}i_{2}}|^{2}i_{1}(i_{1}+1)\right]^{1/2}.
\end{eqnarray}
Similarly, the constrained values of $\langle I_{2}\rangle_{n}$ and $\langle I_{0}\rangle_{n}$ are calculated.

\begin{figure}
\centering
		\includegraphics[width=0.75\textwidth]{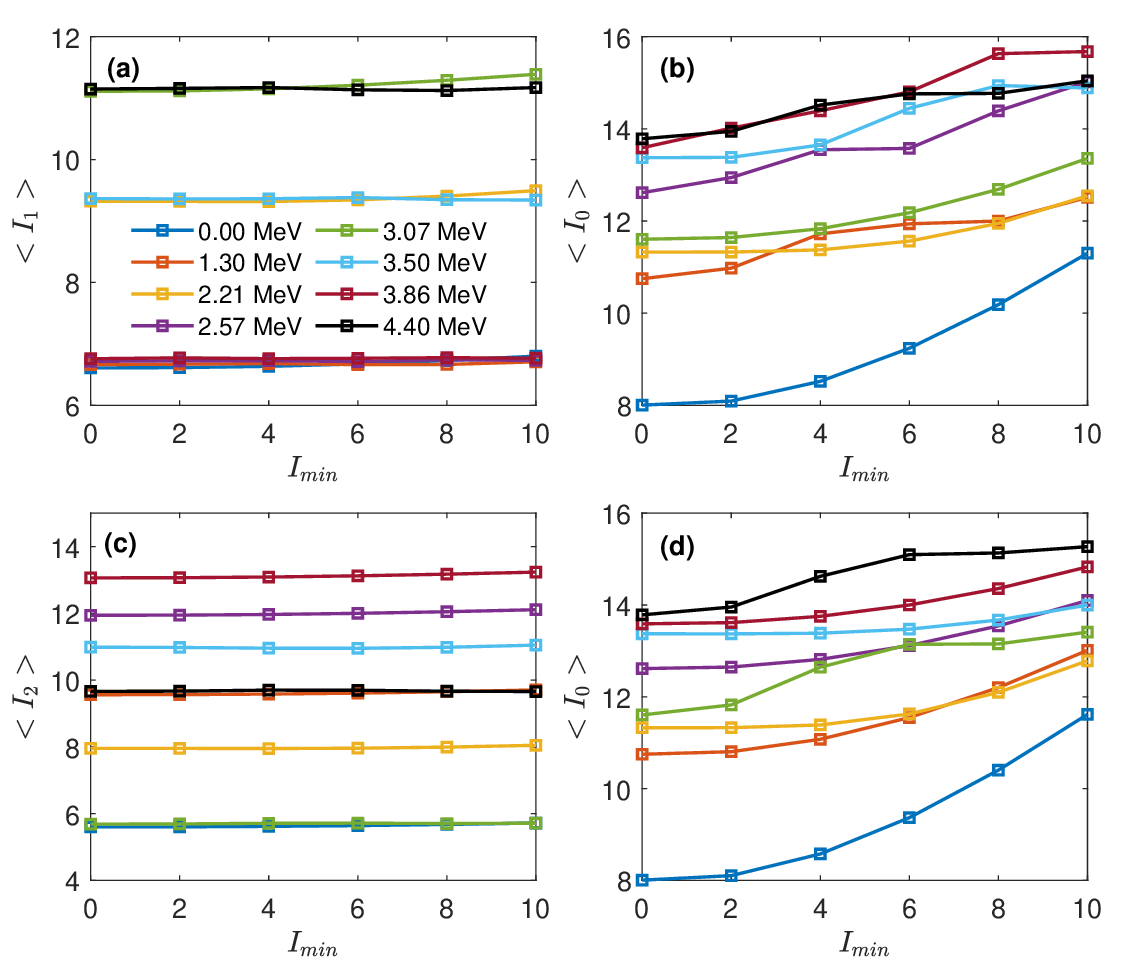}
  \caption{(a,b): The values of $\langle I_{1,0}\rangle$ for the lowest states of angular vibrations with indicated energies in the $^{102}$Zr$(\beta=0.7)$+$^{150}$Ce$(\beta=0.24)$ system with various constraints on the angular momentum of the second fragment $I_{2}\geqslant I_{min}$. (c,d): The same, but for $\langle I_{2,0}\rangle$ with constraints $I_{1}\geqslant I_{min}$.}
  \label{MeanI}
\end{figure}

Figure~\ref{MeanI} display the values of $\langle I_{1,0}\rangle$ for the indicated lowest states of the DNS $^{102}$Zr$(\beta=0.7)$+$^{150}$Ce$(\beta=0.24)$ calculated with various constraints on $I_{2}$ (panels (a,b)), and of $\langle I_{2,0}\rangle$ with various constraints on $I_{1}$ (panels (c,d)).
As observed,  the angular momenta of the first and second fragments are largely independent for all energy levels. However, any constraint on the angular momentum of the fragments strongly affects $\langle I_{0} \rangle$. This implies that any vibration of individual fragments is compensated by the rotation of binary system as a whole, aligning with the large moment of inertia of the relative rotation compared to the individual fragments.

\begin{figure}
\centering
		\includegraphics[width=0.5\textwidth]{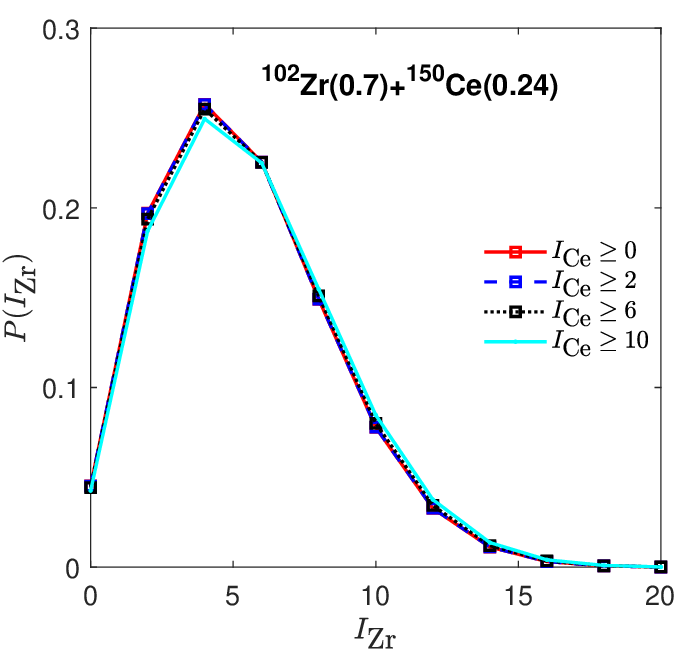}
  \caption{The probability distribution $\sum_{{I_0,I_{Ce}\geqslant I_{min}}}|a_{I_{0},I_{Zr},I_{Ce}}^{(0)}|^2$ of angular momentum of $^{102}$Zr for the ground state of angular vibrations in the $^{102}$Zr$(\beta=0.7)$+$^{150}$Ce$(\beta=0.24)$ system.
  The calculation is performed with various constraints on the angular momentum of $^{150}$Ce. For each constraint, the probability distribution is normalized to unity.}
  \label{Pdistribution}
\end{figure}

The absence of correlations becomes even more distinct if one plots the probability distribution of $I_{1}$ for different constraints imposed on the angular momentum of the second fragment (Fig.~\ref{Pdistribution}). Each curve in Fig.~\ref{Pdistribution} is normalized to unity for better clarity. It is evident that the corresponding curves are to large extent identical, indicating the independence of  angular momenta of two fragments.  The observations, including the excitation spectrum, distribution of $K$ values, and the absence of correlations between the fragments, support the idea of nearly independent angular vibrations of fragments around a pole-to-pole configuration.  This model also suggests that the angular momenta of fragments increase with their deformations that explains the experimentally observed saw-tooth behavior in angular momentum versus fragment mass \cite{Wilson2021}.

An increase in deformation of one fragment results in an increase of its average angular momentum, while the angular momentum of the partner fragment remains almost unchanged. This relationship holds true for all states shown in Fig.~\ref{Spec}, including the ground state.  Indeed, the average distortion angles of the fragments can be estimated as $\overline{\varepsilon_i} = \sqrt{\langle \varepsilon_i^2\rangle }\sim \hbar (C_i \Im_i)^{-1/4}$. From the uncertainty relation, we estimate the average angular momentum of the fragment as $\sim(C_i \Im_i)^{1/4}$. Using Eqs.~(\ref{eq11b})-(\ref{eq8}), one can estimate the stiffness parameter for small amplitude angular vibrations of the fragment as $C_{i}\sim Z_{1}Z_{2}\beta_{i}/R_{m}^{3}(0,0)$. Thus, an increase in deformation of a particular fragment, leads to an increase in both its moment of inertia $\Im_i$ and the corresponding stiffness parameter $C_i$, ultimately resulting in a larger angular momentum. The orientation pumping mechanism of angular momentum generation which arises from uncertainty principle was previously considered in Ref.~\cite{Mikhailov1999}.

\begin{figure}
\centering
		\includegraphics[width=0.5\textwidth]{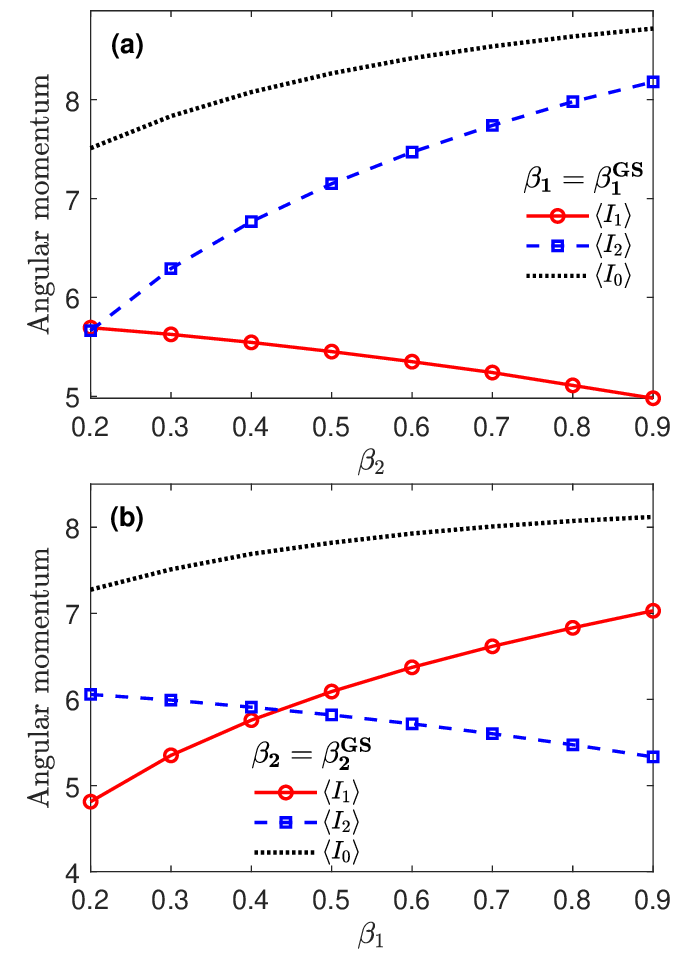}
  \caption{Average angular momenta $\langle I_{0}\rangle$, $\langle I_{1}\rangle$ and $\langle I_{2}\rangle$ for $^{102}$Zr$(\beta_{1})$+$^{150}$Ce$(\beta_{2})$ as a function of deformation $\beta_{2}$ at $\beta_{1}=\beta_{1}^{GS}$ (a), and as a
  function of deformation $\beta_{1}$ at $\beta_{2}=\beta_{2}^{GS}$ (b).}
  \label{BetaDependence}
\end{figure}

To  investigate this quantitatively, we calculate the average angular momenta of the fragments as functions of deformation (Fig.~\ref{BetaDependence} (a,b)).
Each panel shows values of $\langle I_{0,1,2}\rangle$ when the deformation of one fragment corresponds to its ground state, while the deformation of other fragment changes. In agreement with the qualitative estimations given above, the change in the deformation of one fragment corresponds to an increase in its angular momentum, while the average angular momentum of the second fragment remains nearly constant.

The deformations of fragments at scission are necessarily connected with the number of emitted neutrons and corresponding total kinetic energies (TKE). To describe simultaneously the neutron multiplicity and TKE, one has to assume that in the case of large number of neutrons emitted, the fragment deformation should be strongly distorted from its ground state value \cite{Pasca2021,Isaev2023}.

\begin{figure}
\centering
		\includegraphics[width=0.6\textwidth]{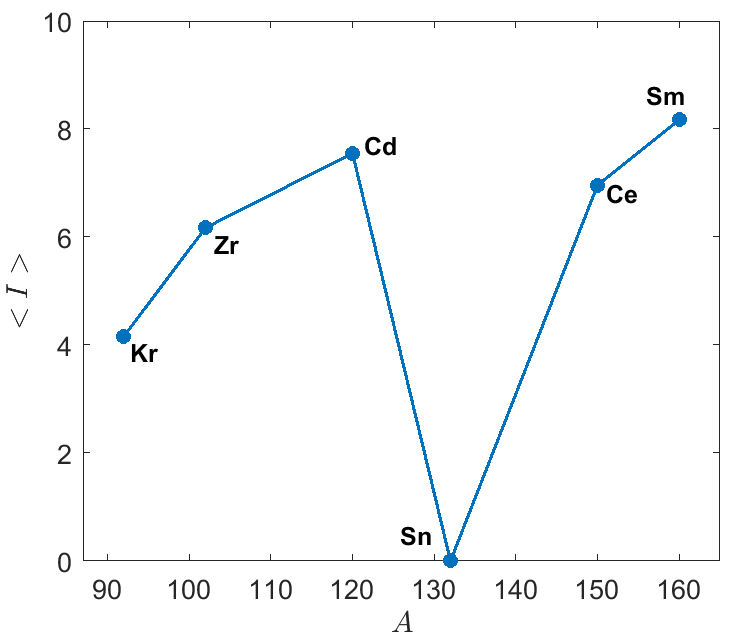}
  \caption{Average angular momentum as a function of fragment mass calculated for the $^{92}$Kr($\beta=0.24$)+$^{160}$Sm($\beta=0.8$), $^{102}$Zn($\beta=0.6$)+$^{150}$Ce($\beta=0.5$), and
 $^{120}$Cd($\beta=0.8$)+$^{132}$Sn($\beta=0$) systems resulting from spontaneous fission of $^{252}$Cf. The calculated points are connected by line to illustrate the emergence of saw-tooth behavior.}
  \label{sawtooth}
\end{figure}

As shown experimentally for $^{252}$Cf \cite{Gook2014}, the increase in number of emitted neutrons is associated with smaller TKE which can only be understood by large deformation of the fragments. Since the neutron multiplicity versus fragment mass distribution exhibits saw-tooth behavior, the same pattern one expects for the deformation distribution. This in turn leads to saw-tooth behavior of the average angular momentum versus fragment masses.
This is illustrated in Fig.~\ref{sawtooth}. The deformations of the fragments in Fig.~\ref{sawtooth} are selected based on an analysis of the relationship between fragment deformation and the experimental average number of neutrons emitted from the fragments \cite{Gook2014}.
In strongly asymmetric splittings, the lighter fragment retains a deformation close to its ground state, while the heavier fragment undergoes significant distortion (see $^{92}$Kr+$^{160}$Sm in Fig.~\ref{sawtooth}). As a result, the lighter fragment has a smaller average angular momentum, whereas the heavier fragment has a relatively larger angular momentum and emits more neutrons due to the larger deformation energy stored. In more symmetric splittings, both fragments exhibit deformations  deviating from their ground states, but the distortion is less pronounced for the heavier fragment compared to asymmetric splittings (see $^{102}$Zn+$^{150}$Ce in Fig.~\ref{sawtooth}). Consequently, this leads to similar values of angular momentum for both fragments. For the mass splittings involving nuclei near Sn, the increased stiffness  with respect to the distortion results in a smaller deformation and, thus, in smaller number of neutrons emitted from the heavier fragment (see $^{120}$Cd+$^{132}$Sn in Fig.~\ref{sawtooth}). This leads to the appearance of saw-tooth behavior of both neutron multiplicity and angular momentum as a function of fragment mass.

Our analysis provides a new explanation of the correlation between the average angular momentum of  fragment and its mass. Unlike the interpretation in Ref.~\cite{Dossing2024}, which attributes the increase in angular momentum to the  increase of excitation energy while assuming the fragment deformation remains close to the ground state, our findings suggest otherwise. We propose that most of the energy available at scission is stored into the  deformation energy. In this case, the angular vibrations in the DNS  are mainly affected by  the lowest states. Thus, the correlation between angular momentum and fragment mass is related to the changes of deformation.


To summarize, we proposed a model for quantum mechanical treatment of angular motion in the  system consisting of two fission fragments at touching. As shown, unlike the bending/wriggling scenario, this motion can be considered as small-amplitude independent vibrations of fragments around pole-to-pole configuration. Angular momentum generated by the vibrations is balanced  by the rotation of the DNS as a whole. The model allows us to explain the absence of correlation between the angular momenta generated in fission fragments. The distribution of projection of angular momenta of the fragments on the separation axis is peaked  around either $K=0$ or $K=1$, but it extends well to $K \approx 4$. The model explains the  saw-tooth pattern in the angular momentum as a function of fragment mass.

\end{document}